# Advances in High Energy Physics

**Role of Doping Ratio on The Sensing Properties of ZnO:SnO$_2$ Thin Films**

Sahar M.Naif , Bushra A.Hasan

*Physics Department, Baghdad University,College of Science, Baghdad ,Iraq*
*Physics Department, Baghdad University,College of Science, Baghdad ,Iraq*

Correspondence should be addressed to sahar M.naif Saharmohammed212@gmail.com

,bushra_abhasan@yahoo.com

## Abstract

Thin films of ZnO:SnO$_2$ were deposited on different substrates like glass and c-Si using spray pyrolysis method .The structures and morphology of the prepared samples films were cheeked using X-ray diffraction and atomic force microscope. Gas sensing measurements provided from resistance measurement in the absent and exposure to NO$_2$ gas . The results showed that good enhancement of sensitivity take place after doping with tin oxide. Maximum sensitivity obtained at 9% doping ratio and operating temperature 200ºC.

## Introduction

Metal oxides, ZnO ,Sno$_2$,Tio$_2$,In$_2$O$_3$, CdO are wide-bandgap *n*-type semiconductors and the most frequently used as a sensitive material for the gas sensors. They have a place with a class of straightforward conductive oxides because of various one of a kind utilitarian properties, of which the most essential are the electrical conductivity, the perceivability in a wide unearthly range and the high reactivity of the surface [1-3].

Metal oxides based gas sensors are generally utilized because of their high sensitivity to harmful for human health or hazardous gases (such as CO, NO,

NO$_2$, H$_2$, etc.) in conjunction with easy fabrication strategies and low assembling costs. Tin (IV) oxide is the most encouraging sensor material among a wide arrangement of semiconducting metal oxides [4-7].

## Materials and Methods

Preparation of an aqueous solutions of tin chloride SnCl2.2H2O and zinc chloride with concentration of 0.1M at (3%,5%,7%,9%) ratios by dissolving in refined water and blended with an attractive stirrer for 15 minute. The splashing mechanical assembly was produced locally in the office labs.
In this technique, the prepared aqueous solutions were atomized by a special nozzle glass sprayer at heated collector glass fixed at thermostatic controlled hotplate heater. Air was used as a bearer gas to atomize the shower arrangement with the assistance of an air blower with weight (7 Bar) wind stream rate (8 cm3/sec) at room temperature. The glass substrate were cleaned with distilled water followed by ethanol in ultrasonically cleaner and then dried.
The temperature was kept up at 250 C amid splashing , The separation between the substrates and shower spout was kept at (30 ±1 cm), number of showering (100) and time between two showering (10 sec). The X-beam diffraction (XRD) information of the readied films were taken utilizing CuKα with radiation of wavelength λ =1.5406Å , current =15mA , voltage = 30 kV and scanning speed = 2 deg /min) over the diffraction angle range 2θ=20-80 at room temperature . The normal crystallite size (D) was assessed utilizing the Scherrer condition as follows[8] :

$D = 0.9\lambda/\beta \cos\theta$ …………………. (1)

where λ, β, and θ are the x-ray wavelength, the full width at half half most extreme (FWHM) of the diffraction pinnacle and Bragg's diffraction angle respectively. The surface distribution of ZnO: SnO$_2$ thin films were measured using a scanning probe microscopy (CSPM5000) instrument. D.C conductivity was calculated using equation (2)

$$\sigma_{D.C} = \frac{1}{\rho} \quad \text{…… …(2)}$$

σ $_{d.c,}$ : The conductivity of the films , ρ: The resistivity of the films

Activation energies can be calculated from the plot of ln σ versus 1000/T according to equation [9]:

$$\sigma = \sigma_0 \exp(\frac{-E_a}{k_B T}) \quad \ldots\ldots(3)$$

where $\sigma_o$ is the base electrical conductivity at 0 °K, Ea is the enactment vitality which compares to (Eg/2) for inborn conduction, T is the absolute temperature and kB is Boltzman's constant equal to $(8.617 \times 10^{-5}$ eVK$^{-1})$ . By taking (Ln) of the two sides of equation (3):

Lnσ = Ln σ$_0$ (-E$_a$ /k$_B$T) …………………..…(4)

From the slope of ln σ against 1000/T graph , the activation energy can be found:

E$_a$ = k$_B$ . slope …………………………..… (5)

Hall measurements are used to distinguish whether a semiconductor is n or p – type

From the Hall coefficient equation, the carrier's concentration of the semiconductor, and the carrier type, can be determined from the sign of $R_H$ such that:

$$R_H = \frac{-1}{n.q} \quad \text{For} \quad \text{n-type} \ldots\ldots\ldots\ldots\ldots.. (6)$$

$$R_H = \frac{1}{p.q} \quad \text{For} \quad \text{p-type} \ldots\ldots\ldots\ldots\ldots\ldots(7)$$

Where $R_H$: Hall coefficient.

The Structure and Mechanism of ZnO for Gas Sensing Behavior there are two essential capacities which a gas sensor comprise of. These are receptor capacities and transducer capacities. Receptor work incorporates the acknowledgment the compound substance, though transducer work changes over the synthetic flag into electrical signals. This area manages the basic properties favoring receptor elements of ZnO in charge of gas detecting conduct.
ZnO has a wide range of auxiliary structures and shapes developed under various development conditions. Wurtzite is the most supported type of ZnO at surrounding conditionsthermodynamically .The lattice constant parameters of wurtzite ZnO are :a=3.2490 Å and c=5.2070 Å with two interconnecting hexagonal close- packed (hcp) sub-lattices in hexagonal lattice of Zn$^2$ and O$^{2-}$ involving sp$^3$ covalent bonding [10] .

## Results and Discussion

The XRD patterns of pure ZnO and doped with SnO$_2$ in different ratios (3%,5%,7%,9%) are shown in Figure(1).

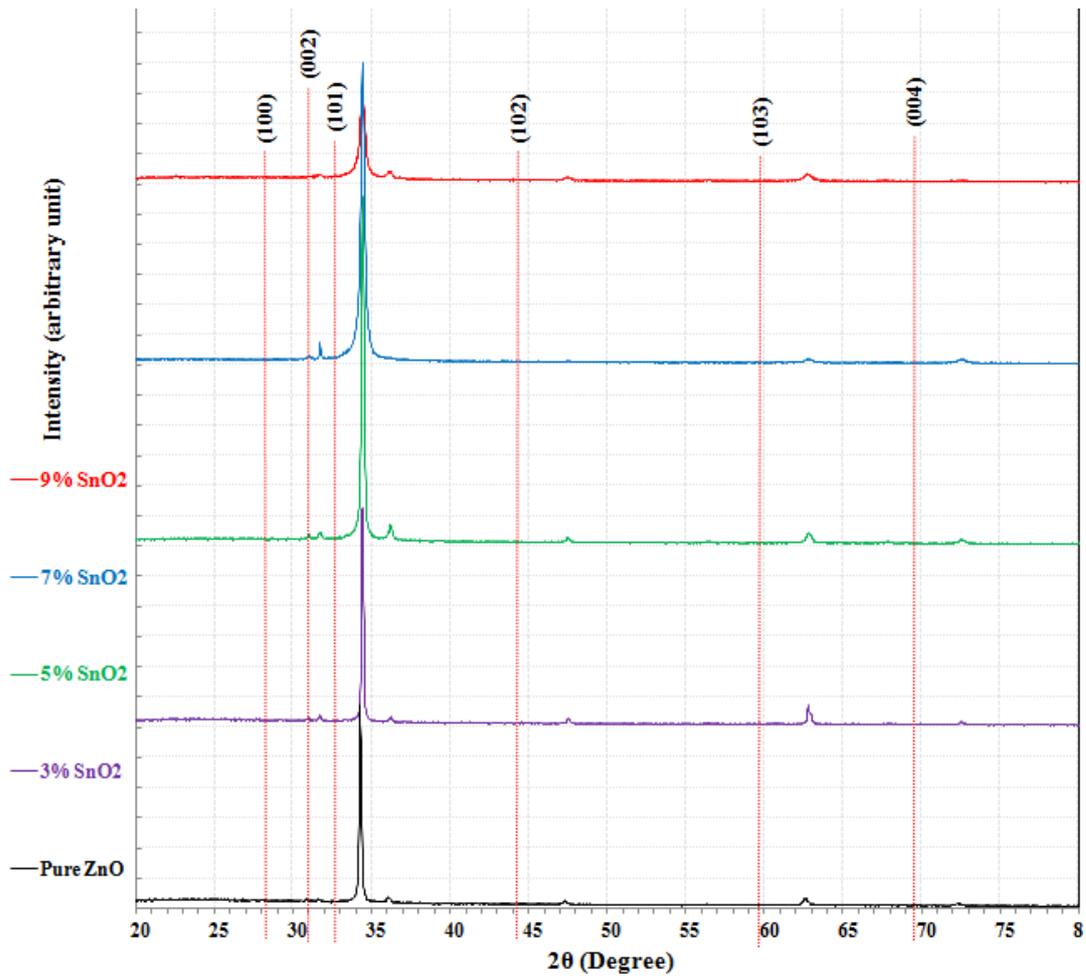

Figure.1 :XRD patterns of ZnO: SnO₂ thin films deposited with different doping ratios.

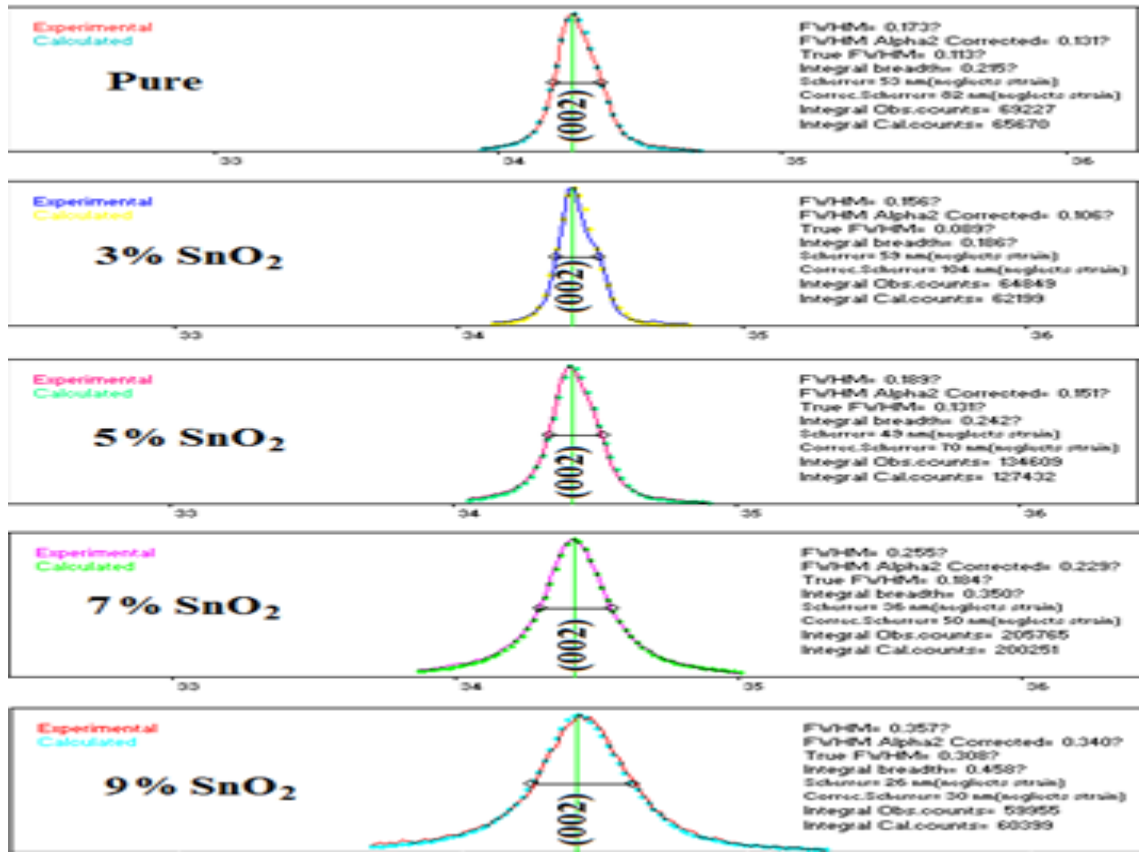

Figure .2 : Variety of FWHM with expanding of SnO2 of ZnO:SnO2 thin movies at favored (002) plane.

Table .1 The XRD out put of ZnO:SnO2 thin movies kept with various doping proportions.
.

| $SnO_2$% | 2θ (Deg.) | FWHM (Deg.) | $d_{hkl}$ Exp.(Å) | D (nm) | (hkl) | $d_{hkl}$ Std.(Å) |
|---|---|---|---|---|---|---|
| 0 | 34.2938 | 0.1730 | 2.6128 | 48.1 | (002) | 2.6035 |
| 3 | 34.4068 | 0.1560 | 2.6044 | 53.3 | (002) | 2.6035 |
| 5 | 34.4633 | 0.1890 | 2.6003 | 44.0 | (002) | 2.6035 |
| 7 | 34.4633 | 0.2550 | 2.6003 | 32.6 | (002) | 2.6035 |
| 9 | 34.4068 | 0.3570 | 2.6044 | 23.3 | (002) | 2.6035 |

Table (1) showed that crystallite size increases as tin oxide introduced to the host material ( zinc oxide). The crystallite measure at that point diminishes because of the widening in the diffraction crests happen because of expanding of pore dividers thickness and upward moves (unwinding of strain) this concur with [11],[12] .

**Atomic force Microscopy (AFM)**

Figure(3)demonstrates the Atomic power microscopy (AFM) pictures for ZnO: SnO2 thin movies with various doping proportions (3%,5%,7%,9%) kept on glass substrate. AFM parameters contain normal measurement, normal unpleasantness and pinnacle top an incentive for these example have been appeared in Table (2) . This table illustrates reduction in average diameter up to 5% followed by increment and finally return to reduced by increasing of doping ratio this agree with [12].

**Table( 2 ) AFM measurements for thin films at different sno$_2$ doping ratios.**

| % SnO$_2$ | Average diameter (nm) | Average roughness (nm) | Peak-peak (nm) |
|---|---|---|---|
| 0 | 91.21 | 6.59 | 37.3 |
| 3 | 88.64 | 4.96 | 20.3 |
| 5 | 74.74 | 10.8 | 48.3 |
| 7 | 105.70 | 9.19 | 61.4 |
| 9 | 99.21 | 8.75 | 43 |

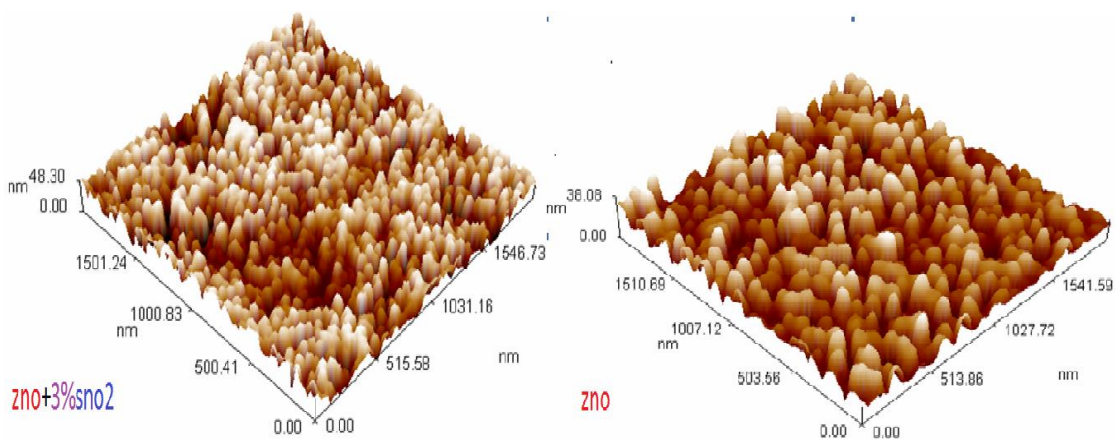

6Hindawi Template version: May18

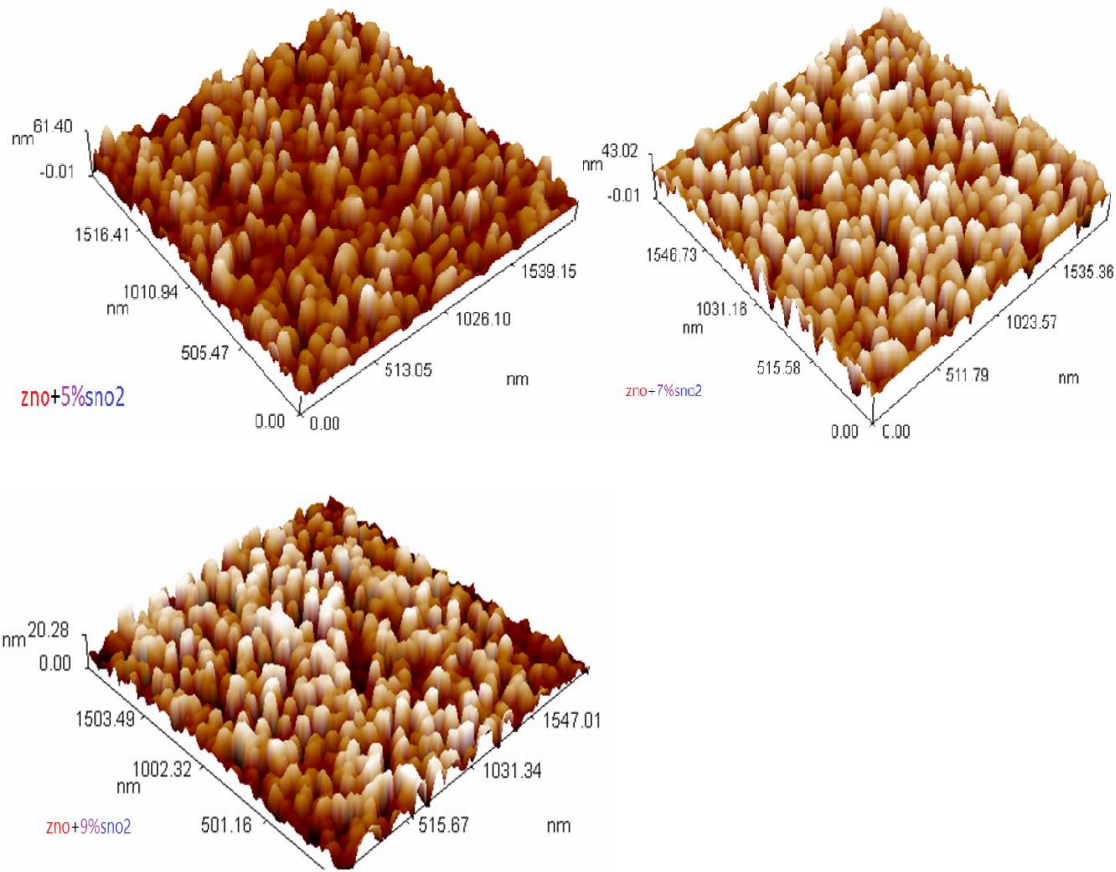

Figure .3:AFM images of ZnO:SnO$_2$ thin films deposited with different sno$_2$ doping ratios.

## The Electrical Measurements

## D.C Conductivity:

Figure(4) demonstrates the variety of Ln(σ) with equal temperature. Clearly there are in excess of one conduction instrument and thus in excess of one actuation vitality. Table (3) appears there are two enactment energies can be watched for the unadulterated and doped examples with,3%,5%,7%,9%. It is clear that the conductivity increased as SnO$_2$ was added to the host material. This result is expected since the increases the density of charge carriers . From other side the activation energy at low temperature denoted by E$_{a2}$ increases by increasing of doping ratio(although the reduction was not in systematic manner). This take place as result of reduction of crystal size as seen from X-ray diffraction and table 1.

Table (3): D.C activation energies, their temperature ranges for pure Zno and doped with different ratio of SnO$_2$ .

| Sample | Activation energy($E_{a1}$)(eV) | Temperature range (K) | Activation energy($E_{a2}$)(eV) | Temperature range (K) | σ $_{R.T}$(ohm.cm)$^{-1}$ |
|---|---|---|---|---|---|
| ZnO | 1.527 | 298-373 | 1.134 | 373-463 | 8.46x10$^{-4}$ |
| ZnO doped with 3% SnO$_2$ | 1.532 | 298-373 | 1.169 | 373-463 | 9.98x10$^{-4}$ |
| ZnO doped with 5% SnO$_2$ | 1.545 | 298-373 | 1.222 | 373-463 | 14.43x10$^{-4}$ |
| ZnO doped with 7% SnO$_2$ | 1.545 | 298-373 | 1.138 | 373-463 | 16.16x10$^{-4}$ |
| ZnO doped with 9% SnO$_2$ | 1.545 | 298-373 | 1.161 | 373-463 | 23.15x10$^{-4}$ |

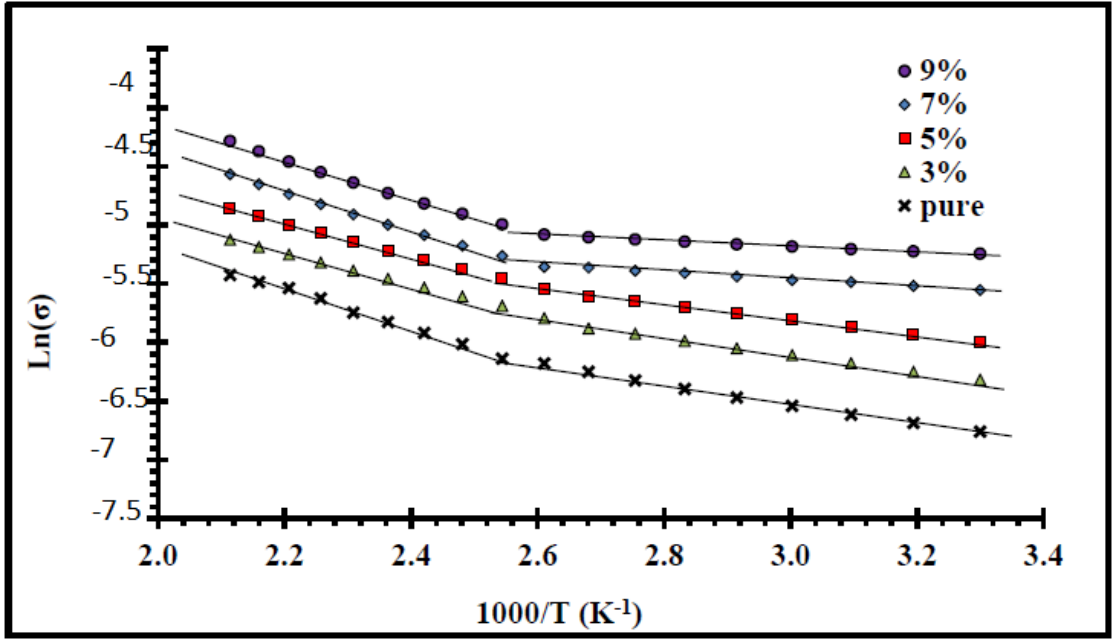

**Figure(4): Plot of ln(σ) vs. 1000/T of un-doped ZnO and doped with different ratio of SnO$_2$.**

**Hall Effect measurements**

Table(4) show the type of charge carriers, concentration (n$_H$) and Hall mobility (μ$_H$), have been estimated from Hall measurements for un-doped ZnO and doped with SnO$_2$ films at different concentration.

**Table(4): Hall measurements results of ZnO :SnO$_2$ thin films at different doping ratios.**

| SnO$_2$(%) | σ$_{R.T}$ (Ω.cm.)$^{-1}$ | R$_H$ (cm$^3$/Coul) | n$_H$ ×10$^{15}$(cm$^3$) | type | μ$_H$(cm$^2$/v.sec) |
|---|---|---|---|---|---|
| 0 | 2.168*10$^{-4}$ | -3859 | 4.402*10$^8$ | N | 0.8366 |
| 3 | 1.778*10$^{-3}$ | -1258 | 4.961*10$^8$ | N | 2.2367 |
| 5 | 1.868*10$^{-2}$ | -1018 | 1.617*10$^9$ | N | 19.016 |
| 7 | 1.252*10$^{-1}$ | -1007 | 1.833*10$^{11}$ | N | 126.07 |
| 9 | 2.1 | -988 | 5.744*10$^{13}$ | N | 2074.8 |

Hall measurements show that all these films have a negative Hall coefficient (n-type charge carriers), this is attributed to following two reasons .The number of electrons excited above the conduction band mobility edge is larger than the number of holes excited below the valance band mobility edge. The life time of free electrons excited from negative defect state is higher than the life time of free holes excited from positive defect state.The charge carriers density increases five order of magnitude by increasing of doping ratio.

## Gas Sensing Measurement
## NO₂ Sensing Mechanism for ZnO:SnO₂ Films

The thin films specimens are examined for gas sensing using NO₂ with concentration of 25 ppm at different operation temperature beginning from room temperature (30°C) up to 200°C with step of 50°C.

Figures from (5) to (9) show the variation of resistance as a function of time with on/off gas valve.

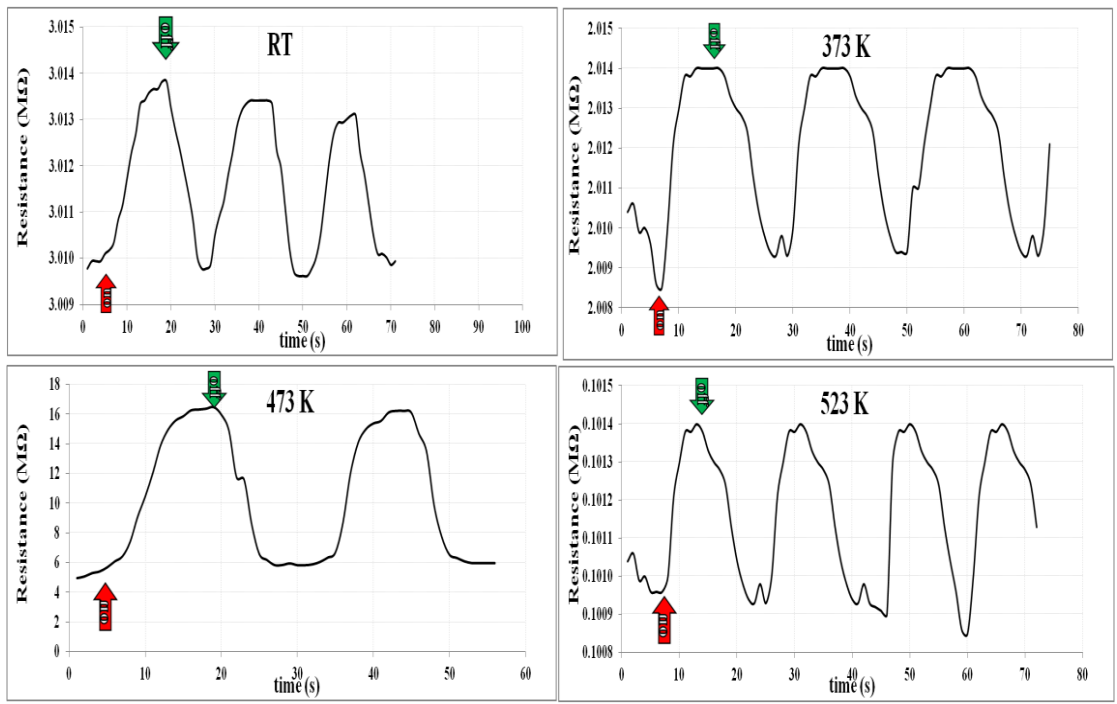

**Figure.5: Variation of resistance with time for Pure films as NO₂ gas**.

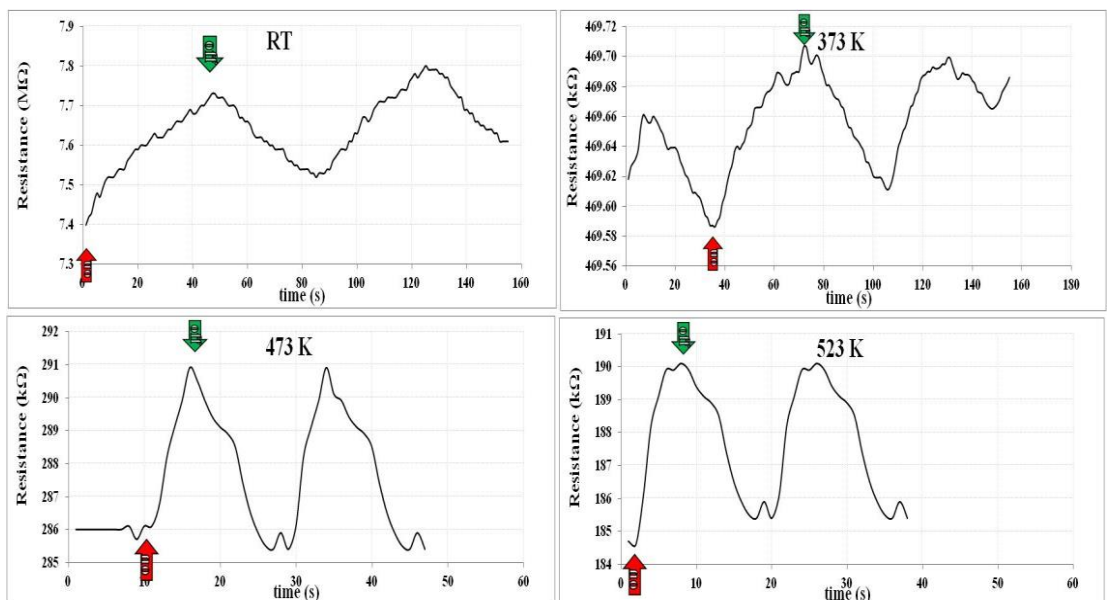

**Figure. 6: Variation of resistance with time for 3% SnO₂ doped ZnO films as NO₂ gas sensing.**

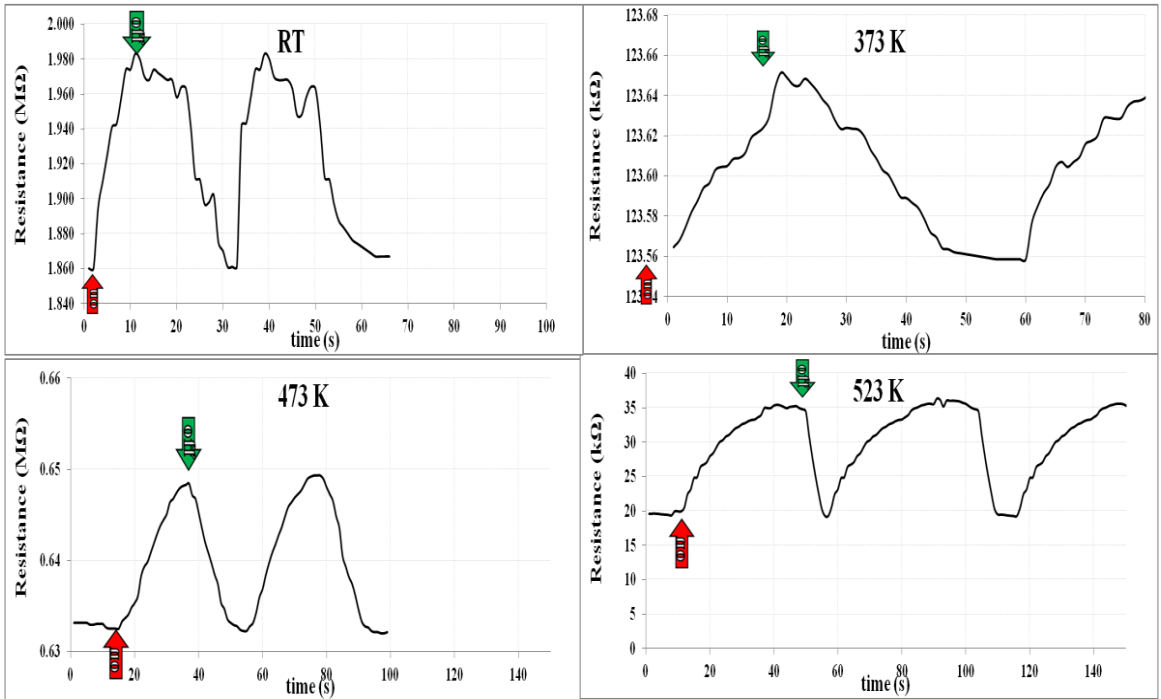

**Figure.7: Variation of resistance with time for 5%SnO$_2$ doped ZnO films as NO$_2$ gas sensor.**

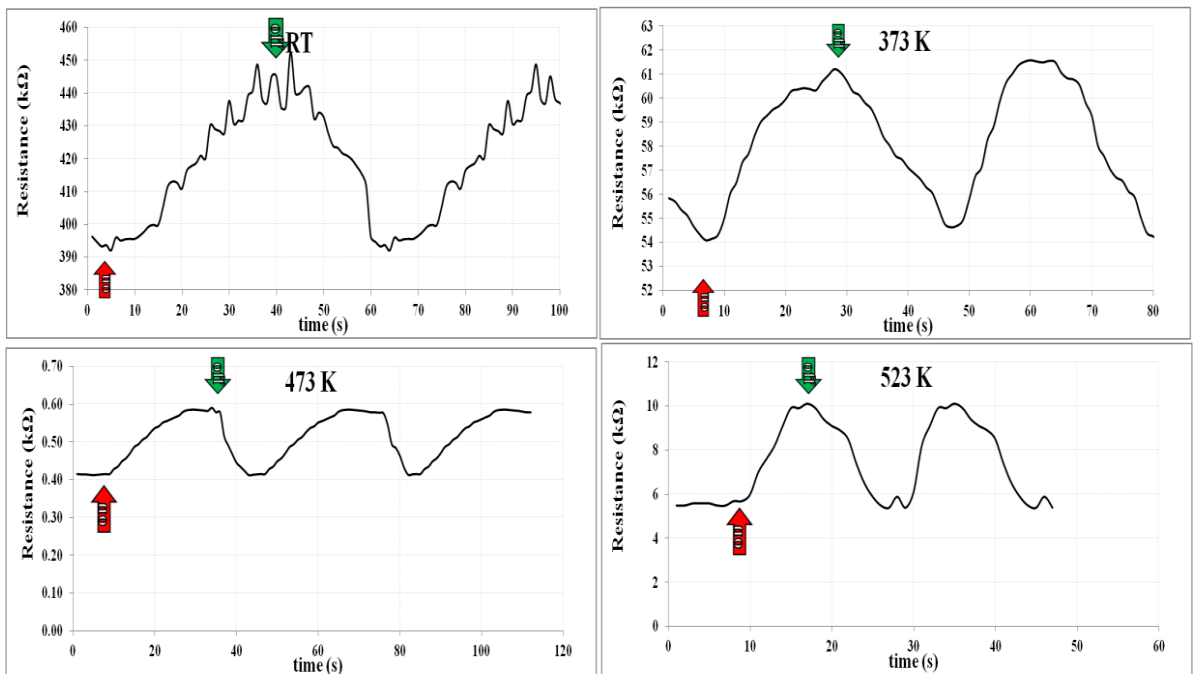

**Figure.8: Variation of resistance with time for 7% SnO$_2$ doped ZnO films as NO$_2$ gas sensor.**

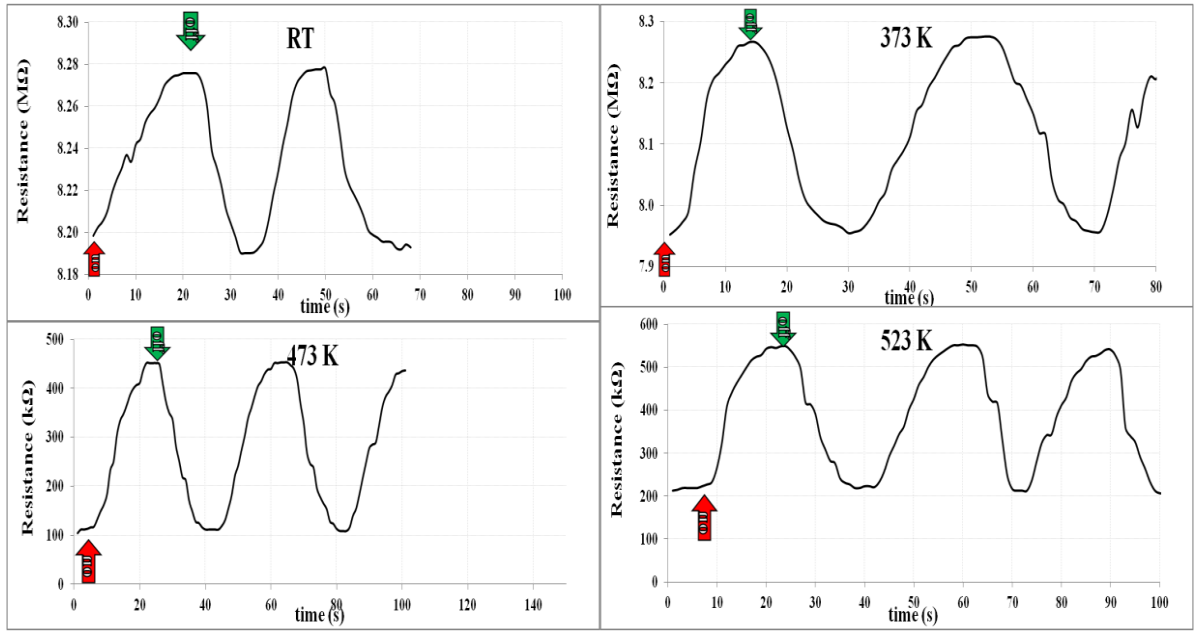

**Figure./9: Variation of resistance with time for 9% SnO$_2$ doped ZnO films as NO$_2$ gas sensor.**

**Table.5:Sensitivity as a function of operating temperature to NO$_2$ gas for pure ZnO and doped with SnO$_2$ at different concentrations deposited on n-Si**

| Operating Temp.(°C) | Sensitivity | | | | |
| --- | --- | --- | --- | --- | --- |
| | doping ratio of SnO$_2$ | | | | |
| | 0% | 3% | 5% | 7% | 9% |
| R.T | 5.9 | 2.7 | 6.45 | 21.15 | 5.97 |
| 100 | 6.255 | 5.988 | 89.0 | 153.8 | 50.6 |
| 200 | 39.2 | 123.7 | 42.6 | 220 | 350 |
| 250 | 29.8 | 59.5 | 75 | 237.5 | 280 |

The results show that the sensitivity of the gas sensors thin films with doping concentrations( 0,3%,5%,7% and 9%) increases with the increase of the operating temperature . It was found(at low working temperature to be specific R.T and 100 oC) there was an expansion in affectability with increment of the of doping fixation up 7% and after that arrival to diminish at high doping proportion low working temperature, while at high working temperature 200 and 250 oC there was a dynamic increment in affectability by expanding of doping proportion this concur with [13].

**Conclusion**

1- Crystal size increases by increasing of doping ratio.
2-Doping has no effect on the type of conductance of the prepared films.
3-Doping ratio has no effect on the high temperature activation energy while has well pronounced effect on the law temperature activation energy 4-Maximum sensitivity obtained at the lowest crystal size belonged to ZnO thin films doped with 9% $SnO_2$ at optimal temperature of (473K).

**Data Availability**

The data used to support the findings of this study are available from the corresponding author upon request.

**Disclosure**

Sahar M.naif is co-first author

**Conflicts of Interest**

The author declares that there are no conflicts of interest regarding the publication of this paper.

**References**

[1] M. Batzill "Surface science studies of gas sensing materials: SnO2. Sensors" 6: 1345-1366.2006.

[2] C.Wang , L.Yin , L.Zhang , D.Xiang ,R. Gao  Metal oxide gas sensors: Sensitivity and influencing factors. Sensors 10: 2088-2106,(2010).

[3] S. Nagirnyak ,V. Lutz , T.Dontsova , Astrelin I The effect of the synthesis conditions on morphology of tin (IV) oxide obtained by vapor transport method. Springer Proc Phys 183: 331-341,(2016).

[4] J. Pan , H.Shen , S.Mathur  One dimensional SnO2 nanosructures:Synthesis and application.1-12,(2012).

[5] A. Köck , A.Tischner ,T. Maier , M.Kast , Edtmaier C, et al. Atmospheric pressure fabrication of SnO2-nanowires for highly sensitive CO and CH4 detection. Sensors and Actuators B: 160-167,(2009).

[6] JH. Park , JH.Lee  Gas sensing characteristics of polycrastalline SnO2 nanowires prepared by polyol method. Sensors and Actuators B: 151-157,(2009).

[7] L. Qin , Xu J, X.Dong , O.Pan ,Z. Cheng , et. al. The template-free synthesis of square-shaped SnO2 nanowires: The temperature effect and acetone gas sensors. Nanotechnology 19: 1-8,(2008).

[8] J. Q. Xu, Q. Y. Pan, Y. A. Shun and Z. Z. Tian, "Grain Size Control and Gas Sensing Properties of ZnO Gas Sensor," Sensors and Actuators B: Chemical, Vol. 66, No. 1-3,  pp. 277-279,July 2007.

[9] P. Kireev, "Semiconductor Physics " ,Moscow: MIR publishers, (1978).

 [1o]. R. Kumar, G. Kumar, A. Umar," Zinc oxide nanomaterials for photocatalytic degradation of methyl orange": a review. Nanosci. Nanotechnol. Lett. 6(8), 631–650 doi:10.1166/nnl.2014.1879, (2014).

   [11] E. Muchuweni, T.S.Sathiaraj,H.Nyakotyo"Synthesis andcharacterization of zincoxidethinfilmsfor optoelectronic applications": Heliyon3(2017)e00285.doi: 10.1016/j.heliyon.2017.e00285.

[12] S.khafory , N.Talib and M. HasanSuhail"Some Physical Properties of ZnO/SnO2 Thin Films Prepared By Spray Pyrolysis Technique": IOSR Journal of Applied Physics (IOSR-JAP).Volume 8, Issue 5 Ver. I (Sep - Oct. 2016), PP 10-17.

[13] A. Z. Sadek, W. Wlodarski, K. Kalantar-zadeh"ZnO Nanobelt Based Conductometric H2 and NO2Gas Sensors"Sensor Technology Laboratory, School of Electrical &Computer Engineering, RMIT University,Melbourne, Australia. 0-7803-9056.2005.

13